\documentstyle[12pt,aps,eqsecnum,preprint,psfig]{revtex}

\tightenlines

\global\firstfigfalse

\begin{document}

\draft

\title{Synchrotron radiation of crystallized beams}
\author{Harel Primack\footnote{
          email: harel@phyc1.physik.uni-freiburg.de}
        and
        Reinhold Bl\"umel\footnote{
          email: blumel@phyc1.physik.uni-freiburg.de}}
\address{Fakult\"at f\"ur Physik, 
         Albert--Ludwigs Universit\"at Freiburg,
         Hermann--Herder Str.\ 3, 
         D-79104 Freiburg, Germany}
\date{Submitted to Phys.\ Rev.\ E, 29 October 1998}
\maketitle

\begin{abstract}

We study the modifications of synchrotron radiation of charges in a
storage ring as they are cooled. The pair correlation lengths between
the charges are manifest in the synchrotron radiation and coherence
effects exist for wavelengths longer than the coherence lengths
between the charges. Therefore the synchrotron radiation can be used
as a diagnostic tool to determine the state (gas, liquid, crystal) of
the charged plasma in the storage ring. We show also that the total
power of the synchrotron radiation is enormously reduced for
crystallized beams. This opens the possibility of accelerating
particles to ultra-relativistic energies using small--sized cyclic
accelerators.

\end{abstract}

\pacs{29.20.c, 29.27.a, 41.75.i, 41.60.Ap}

\section{Introduction}
\label{sec:intro}

Ion--beam crystallization is an exciting and relatively new field of
physics in which a new state of matter is sought for. Namely, ions
which rapidly circulate in a storage ring and are cooled are expected
to form geometrically--ordered structures (crystals) which have a
density much smaller than normal crystalline solids \cite{SK85,HG95}.
Although great effort is currently invested in achieving such crystals
\cite{HG95,Ste96,Han91}, and cooling techniques were significantly 
improved \cite{Lau98}, there is still no clear--cut experimental
evidence for them. It is hoped, however, that crystalline beams will
be produced in the near future.

Synchrotron radiation, on the other hand, is a very well-established
field of physics that has been investigated continuously from the
early days of particle accelerators. Many synchrotron sources are
operating around the world (e.g., DESY (Hamburg, Germany), NSLS
(Brookhaven, USA), KEK (Tsukuba, Japan)), and many applications
already exist \cite{HHK94}.

It is the purpose of this paper to establish a link between beam
crystallization and synchrotron radiation. This link is two--fold:
\begin{itemize}

  \item To use synchrotron radiation and modifications thereof in
  order to {\em detect} the creation and existence of beam
  crystals. This is required since for fast beams direct detection
  methods are difficult to implement \cite{HG95}. Thus synchrotron
  radiation can be used as an indirect diagnostic method to detect the
  formation of beam crystals. The diagnostic methods discussed below
  are also applicable to liquid and gaseous beams.

  \item Even more importantly, once beam crystals are formed, they can
  be used to {\em modify}\/ the synchrotron radiation with respect to
  the ordinary incoherent case. Therefore one can achieve dramatic
  suppression and enhancement effects of synchrotron radiation using
  crystallized beams \cite{PB98}. In particular, the total power that
  is radiated from an equi-spaced circulating chain of particles is
  much smaller (in the appropriate limits \cite{PB98}) than the
  radiation from the same number of randomly--located particles. This
  opens the possibility for accelerating particles to
  ultra-relativistic energies with little radiation loss, which is
  currently the main limitation of circular electron accelerators.
  Thus the suppression of synchrotron radiation by beam
  crystallization may eventually lead to the construction of
  smaller--sized circular electron accelerators.
\end{itemize}

In the following we shall detail the connection between beam crystals
and synchrotron radiation. It is important to emphasize that currently
researchers are trying to obtain beam crystals of heavy ions that can
be cooled with electrons and lasers.  For heavy ions, however, the
synchrotron radiation is small.  Even for protons, e.g., the lightest
of the ``heavy ions'', the synchrotron radiation is about a factor
$10^{-13}$ smaller than for electrons with the same energy.  Thus, we
expect that realistically the effects predicted in this paper will be
important only for liquid or crystallized electron beams. This,
however, poses the challenge of obtaining crystallized electron
beams. Thus we hope that the ideas put forward in this paper will
motivate experimentalists to work towards obtaining crystallized
electron beams.  In any case, however, the analysis presented below
applies to any species of charged particles. Thus the theory can in
principle be verified for ion--beam crystals. Also we stress from the
outset that the effects discussed below go beyond what is known as
``coherent synchrotron radiation'' which is the coherent enhancement
of synchrotron radiation of small electron bunches for wavelengths
that are longer than the bunch size (see, e.g.,
\cite{Mic82,Nak89,BHS91,Kat98} and Sec.\ \ref{sec:bunched} below).

The paper is structured in the following way. In Sec.\
\ref{sec:theory} we discuss the general theory that underlies the 
suppression and enhancement effects of synchrotron radiation. In
Sec.\ \ref{sec:form-factor} the theory is applied to the three phases
of a coasting charged--particle beam that occur in practice: Gaseous,
liquid and crystal. In Sec.\ \ref{sec:bunched} the necessary
modifications of the theory for a bunched beam are discussed. In Sec.\
\ref{sec:total-power} we present analytical and numerical results 
concerning the suppression of synchrotron radiation by a crystalline
beam. Finite temperature effects are discussed explicitly. In Sec.\
\ref{sec:summary} we discuss our results and conclude the paper with
proposals for experimental applications of the effects discussed in
this paper.

\section{General Theory}
\label{sec:theory}

We consider $N$ charged particles with charge $q$ circulating in a
circular storage ring of radius $\rho$ with velocity $v$. The charges
are assumed to be coherent with a reference circulating charge, but
are allowed to have constant time lags $\Delta t_j$ as well as
constant spatial displacements $\Delta {\vec r}_j$ from the reference
orbit. According to the theory of radiation of moving sources, the
total power that is emitted from the $N$ charges is given by
\cite{Sch46,Jac75}
\begin{equation}
  I^{(N)} = 
  \sum_{n=1}^{\infty} g_n I^{(1)}_n,
  \label{eq:totpow}
\end{equation}
where $I^{(1)}_n$ is the power that is emitted with frequency
$\omega_n \equiv n \omega = n v / \rho$ due to a single circulating
particle, and $g_n$ is the form factor of the beam. The explicit
expression for $I^{(1)}_n$ is \cite{LL79}
\begin{equation}
  I^{(1)}_{n} =
  \frac{q^2 c \beta}{2 \pi \epsilon_0 \gamma^2 \rho^2}
    \left[ \beta^2 \gamma^2 n J_{2n}^{\prime}(2 n \beta) -
      n^2 \int_{0}^{\beta} J_{2n}(2 n \xi) {\rm d}\xi
    \right],
 \label{eq:I1nexact}
\end{equation}
where $\beta \equiv v/c$, $c$ is the speed of light, $\gamma \equiv
1/\sqrt{1-\beta^2}$, and $J_n$ are the ordinary Bessel functions
\cite{GR94}. The form factor is given by
\begin{equation}
  g_n = 
  \left| \sum_{j=1}^{N} \exp ( i n \phi_j ) \right|^2 =
    \sum_{j, j'=1}^{N} \exp [ i n ( \phi_j - \phi_{j'} ) ]
  \label{eq:form-factor}
\end{equation}
and the angles $\phi_j$ are given by \cite{KT91}
\begin{equation}
  \phi_j =
  \omega \Delta t_j +
  \frac{\beta}{\rho} \vec{n} \cdot \Delta {\vec r}_j,
\end{equation}
where $\vec{n}$ is the unit vector pointing from the center of the
ring to the observation point. We observe that in the form factor the
role of the time delays and the spatial displacements is equivalent.
Thus we restrict ourselves hereafter to time delays only. This
simplifies the calculations and gives qualitatively the same results.
It is also compatible with the current experimental trend according to
which linear ion crystals (one--dimensional crystals) are sought
for. We shall denote in the following the phase differences by
$\theta_j \equiv \omega \Delta t_j$.

Suppose now that we treat the quantities $\theta_i$ as random
variables distributed according to the normalized probability density
$P(\theta_1, \ldots, \theta_N)$. Then the expectation value of the
total power is
\begin{equation}
  \langle I^{(N)} \rangle =
  \sum_{n=1}^{\infty} \langle g_n \rangle I^{(1)}_n,
  \label{eq:average-power}
\end{equation}
where
\begin{equation}
  \langle g_n \rangle =
  \int_{0}^{2 \pi} \,
    {\rm d}\theta_1 \cdots {\rm d}\theta_N \,
    P(\theta_1, \ldots, \theta_N)
    \sum_{j, j'=1}^{N} \exp [ i n ( \theta_j - \theta_{j'} ) ].
  \label{eq:ff-prob}
\end{equation}
This can be rewritten as
\begin{equation}
  \langle g_n \rangle =
  \int_{- 2 \pi}^{+ 2 \pi} \, {\rm d}\Delta \,
  {\rm e}^{i n \Delta} R_2 (\Delta),
\end{equation}
where
\begin{equation}
 R_2 (\Delta) \equiv
 \int_{0}^{2 \pi} \, {\rm d}\theta_1 \cdots {\rm d}\theta_N \,
    P(\theta_1, \ldots, \theta_N)
    \sum_{j, j'=1}^{N} \delta [ \Delta - ( \theta_j - \theta_{j'} ) ]
  \label{eq:r2-def}
\end{equation}
is the two--point correlation function, i.e.\ the (non-normalized)
chance of finding a pair of $\theta$'s a distance $\Delta$ apart.
Therefore we conclude that the crucial quantity that determines the
modifications of synchrotron radiation due to coherence effects is
$R_2$, our main object of study. The physics of the particle beam
(temperature, structure) is reflected in $R_2$ and is consequently
linked to modifications of the synchrotron radiation.

Before applying the above formulas, we make some further simple
manipulations. To avoid complications with the $2\pi$-periodicity we
define
\begin{equation}
  \hat{R}_2(\Delta) \equiv
  2 [ R_2 (\Delta) + R_2 (2 \pi - \Delta) ].
\end{equation}
Using the relation $R_2 (\Delta) = R_2 (-\Delta)$, easily
derived from (\ref{eq:r2-def}), we obtain
\begin{equation}
  \langle g_n \rangle =
  \int_{0}^{\pi} {\rm d}\Delta \,
    \cos ( n \Delta ) \hat{R}_2(\Delta).
\end{equation}
This can finally be recast as
\begin{equation}
  \langle g_n \rangle =
  N +
  \int_{0}^{\pi} {\rm d}\Delta \,
    \cos ( n \Delta ) \tilde{R}_2(\Delta),
  \label{eq:ff-rtilde}
\end{equation}
where $\tilde{R}_2(\Delta)$, defined in $[0,\pi]$, is the two--point
correlator that does not include the ``diagonal'' part $2 N
\delta(\Delta)$, emerging from the $j=j'$ terms of $R_2$.

An important special case is the case of independent particles, i.e.\
\begin{equation}
  P(\theta_1, \ldots, \theta_N) =
  \prod_{j=1}^{N} P_1 (\theta_j),
  \label{eq:prob-independent}
\end{equation}
where $P_1$ is the (normalized) one--point density of the
particles. For this case the resulting form factor is
\begin{equation}
  \langle g_n \rangle =
  N + N(N-1) \left| \int_{0}^{2 \pi} \, {\rm d}\theta
    P_1(\theta) {\rm e}^{i n \theta} \right|^2.
  \label{eq:gn-independent}
\end{equation}
%

\section{Application to cooled particle beams}
\label{sec:form-factor}

In the following we shall study a few representative situations of a
particle beam as it is being cooled and crystallized. We shall
qualitatively infer the form of the two--point correlator $\tilde R_2$
for each of the cases, and calculate the resulting form factor of the
synchrotron radiation. Thus the focus in this section is on the
spectral modifications of the synchrotron radiation expressed by the
behavior of $\langle g_n \rangle$. We shall show that the
modifications of $\langle g_n \rangle$ as the temperature is lowered
defines an excellent tool for the diagnostics of the thermodynamic
state of the beam. The modifications due to bunching are considered in
Sec.\ \ref{sec:bunched}. The suppression of the total emitted power is
discussed in Sec.\ \ref{sec:total-power}.

We start with a very hot particle beam. In such a case we expect the
particles to be completely independent. Therefore equations
(\ref{eq:prob-independent}) and (\ref{eq:gn-independent}) apply. For a
particle beam that fills the whole ring (coasting beam) we expect on
the basis of symmetry a uniform distribution
\begin{equation}
  P_1(\theta) = \frac{1}{2 \pi}.
\end{equation}
The form factor becomes
\begin{equation}
  \langle g_n \rangle = N \, , \,\,\, n = 1, 2, \ldots \; .
\end{equation}
Thus, for a hot coasting beam,
\begin{equation}
  \langle I^{(N)} \rangle = N I^{(1)} \ .
\end{equation}
This is what we expect from totally incoherent radiation of $N$
particles.

For beams that are bunched, a typical shape is a Gaussian. The
resulting density is
\begin{equation}
  P_1(\theta) =
  \frac{1}{\sqrt{2 \pi \sigma^2}}
  \sum_{m = -\infty}^{+\infty}
    \exp \left[ - \frac{(\theta - \theta_0 + 2 \pi m)^2}{2 \sigma^2}
        \right],
  \label{eq:p1-gaussian}
\end{equation}
where $\theta_0$ is the location of the center of the bunch and
$\sigma$ is its angular width. The summation over $m$ is to ensure the
$2 \pi$-periodicity. For $\theta_0 \gg \sigma$, $2 \pi - \theta_0 \gg
\sigma$ only the $m=0$ component in (\ref{eq:p1-gaussian})
is significant. The resulting form factor is
\begin{equation}
  \langle g_n \rangle = N + N(N-1) \exp ( - n^2 \sigma^2 ).
  \label{eq:ff-gb}
\end{equation}
In the bunched case, therefore, in addition to the incoherent term
$N$, we also obtain a term that represents the coherent synchrotron
radiation for low harmonics $n \lesssim 1/\sigma$. One obtains
qualitatively the same results for other shapes of the bunch
\cite{NS54,Shi91}.

The above results are well-known and form the basis of the field of
``coherent synchrotron radiation'' in which enhancement of the
radiation is predicted \cite{Sch46,NS54,Mic82,KT91} and experimentally
measured \cite{Nak89,BHS91,Kat98} due to the collection of the charges
(electrons) into small bunches. In our case this applies to the
limiting case of a bunched but very hot beam of particles in which the
particles within the bunch are uncorrelated.

In the following we shall introduce the correlations between the
particles as the beam is cooled and use the full expressions
(\ref{eq:average-power})--(\ref{eq:ff-rtilde}) rather than
(\ref{eq:gn-independent}). These correlations are neglected in the
field of coherent synchrotron radiation since the beam is assumed to
be very hot.  But the correlations become more and more important as
the beam is being cooled. To simplify the treatment, we focus in this
section on coasting beams. In Sec.\ \ref{sec:bunched} we introduce the
necessary modifications to describe bunched beams.

\begin{figure}[tp]

  \begin{center}
    \leavevmode
    \psfig{figure=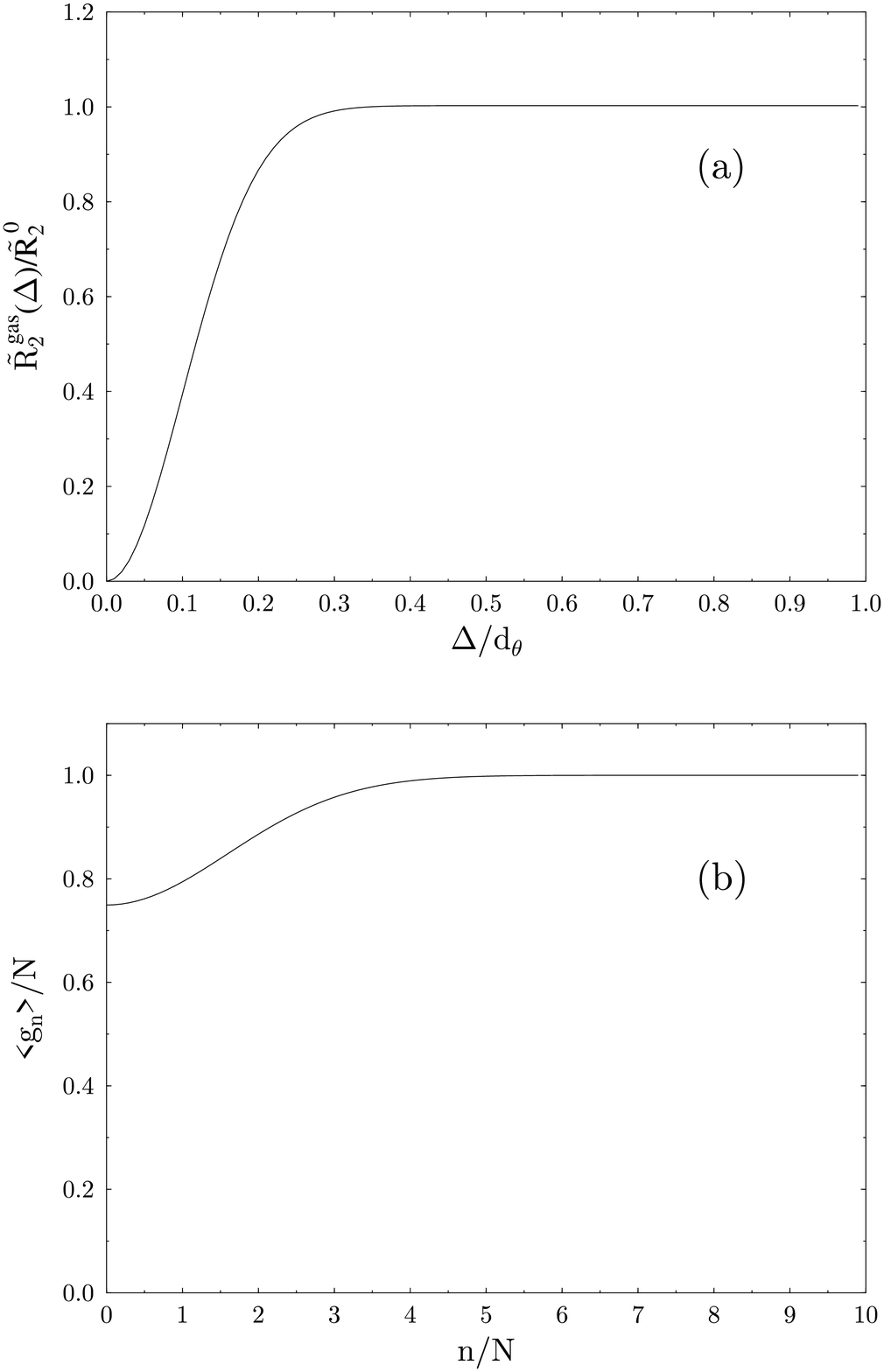,height=18cm}
  \end{center}

  \def\cap:gas{The two--point correlator (a) and the form factor (b)
  for the ``gaseous'' state of the beam, described by
  (\protect\ref{eq:r2-gas}) and (\protect\ref{eq:gn-gas}),
  respectively. We used the parameters $a=d_{\theta}/10, N=100$.}

  \caption{\cap:gas}

  \label{fig:gas}
\end{figure}

If the temperature of the beam is moderately high, we expect that the
particles start to show repulsion from each other. That is, they will
avoid the vicinity of each other due to the Coulomb repulsion and act
like a non-ideal ``gas''. This can be described phenomenologically by
\begin{equation}
  \tilde{R}_2^{\rm gas} (\Delta) =
  c_1 \tilde{R}^{0}_2(\Delta)
  \left[ 1 - \exp \left( - \frac{\Delta^2}{2 a^2} \right) \right],
  \label{eq:r2-gas}
\end{equation}
where
\begin{equation}
  \tilde{R}^{0}_2(\Delta) = \frac{N(N-1)}{\pi}, \ \ \
   0 \leq \Delta \leq \pi
\end{equation}
is the trivial two--point correlator for a uniformly coasting beam of
independent particles and $a$ is the (angular) ``hard--core'' scale of
repulsion. Interpreting (\ref{eq:r2-gas}), we modified
$\tilde{R}^{0}_2$ by a narrow ``dip'' of width $a$ near $\Delta = 0$
(the ``correlation hole'') such that $\tilde{R}_2(0) = 0$ (total
repulsion at $\Delta = 0$). The constant $c_1$ is for
normalization. It is approximately $(1-a/\sqrt{2 \pi})^{-1}$ for small
values of $a$ ($a \ll \pi$). Actually, since we are still in the high
temperature regime, we need to assume that $a \ll 2 \pi / N \equiv
d_{\theta}$, i.e.\ the hard core repulsion occurs on scales smaller
than the mean distance between the particles.  For $N \gg 1$, which is
the interesting case here, the two conditions on $a$ are
consistent. We note that $a$ depends on the temperature and increases
as the temperature decreases. For $a\ll 1$ we obtain the following
form factor:
\begin{equation}
  \langle g_n^{\rm gas} \rangle =
  N - \frac{N(N-1) a}{\sqrt{2 \pi}}
    \exp \left( - \frac{n^2 a^2 }{2} \right) \ ,\ \ \
 n=1,2,\ldots\ .
  \label{eq:gn-gas}
\end{equation}
Since the form factor is a non-negative quantity, we immediately
infer an upper limit on $a$:
\begin{equation}
  a \lesssim \frac{\sqrt{2 \pi}}{N} =
             \frac{d_{\theta}}{\sqrt{2 \pi}}.
\end{equation}
This result is intuitively clear since the hard core cannot be larger
than the mean distance of the particles.  It is also compatible with
the assumptions above concerning $a$. Physically, we observe that
there is a {\em suppression}\/ of the synchrotron radiation for $n
\lesssim 1/a$ (lower harmonics). We can therefore estimate the 
hard--core scale (and hence the temperature) from the coherent
modifications of the synchrotron radiation for the gas--like state of
the particle beam. We note that the overall suppression effect is
small. For small $n$ values, where the reduction of the emitted power
is largest, the relative suppression with respect to the incoherent
case amounts to only
\begin{equation}
\left|
  \frac{\langle g_n^{\rm gas} \rangle - N}{N} \right|
\approx
  \frac{\sqrt{2 \pi} a}{d_{\theta}} \ll 1.
\end{equation}
The situation is illustrated in Fig.\ \ref{fig:gas}.

\begin{figure}[tp]

  \begin{center}
    \leavevmode
    \psfig{figure=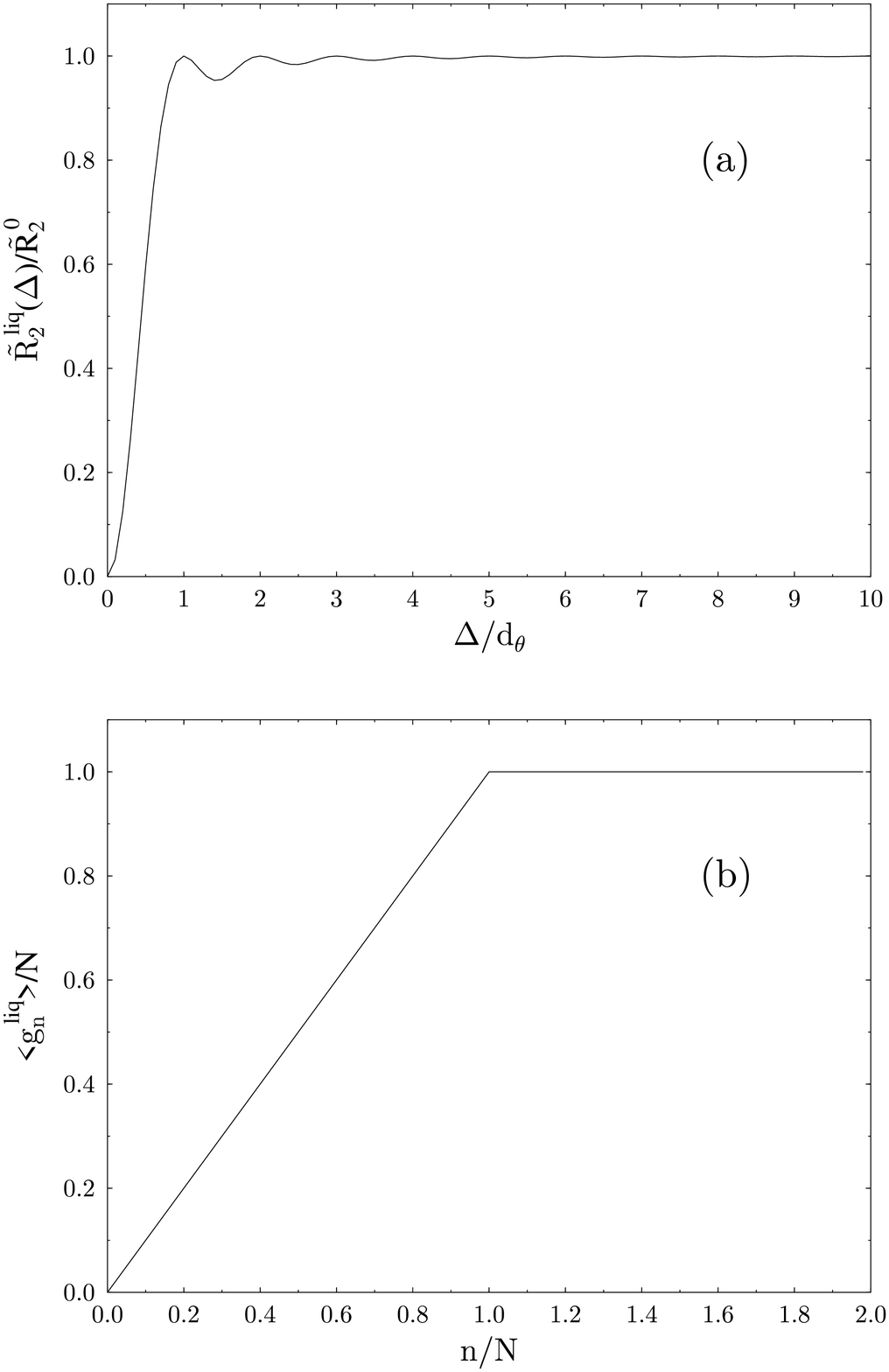,height=18cm}
  \end{center}

  \def\cap:liq{The two--point correlator (a) and the form factor (b)
  for the ``liquid'' state of the beam, described by
  (\protect\ref{eq:r2-liq}) and (\protect\ref{eq:gn-liq}).}

  \caption{\cap:liq}

  \label{fig:liq}
\end{figure}

When the temperature becomes smaller such that the Coulomb energy is
comparable to the thermal energy, we expect the particle beam to
become somewhat ordered and to form a liquid--like plasma. The partial
order is a precursor to crystallization. In particular, the (angular)
distance on which the repulsion between particles is manifest is
$d_{\theta}$, and the order effects should persist over a few mean
distances. The two--point correlator is qualitatively given by
\begin{equation}
  \tilde{R}_2^{\rm liq} (\Delta) =
  c_2 \tilde{R}^{0}_2 (\Delta)
  \left[ 1 - \frac{\sin^2 (\pi \Delta / d_{\theta})}
                  {(\pi \Delta / d_{\theta})^2} \right],
 \label{eq:r2-liq}
\end{equation}
where $c_2 \approx (1 - 1/ N)^{-1}$ is a normalization factor. As
before, we assume $N \gg 1$. The above two--point correlator displays
a strong repulsion for small distances ($\tilde{R}_2^{\rm liq} (0) =
0$) as well as oscillations that persist for a few mean distances. It
eventually reaches the asymptotic limit of uncorrelated particles.
Thus it represents an intermediate situation (``liquid'') between the
slight mutual repulsion (``gas'') treated above and long--range order
(``crystal'') discussed below. The above expression was obtained as a
result of an exact calculation for a one--dimensional chain of
particles with logarithmic repulsion by Dyson in the context of Random
Matrix Theory \cite{Boh89}. The resulting form factor is
\begin{equation}
  \langle g_n^{\rm liq} \rangle = \min(n, N) \, .
  \label{eq:gn-liq}
\end{equation}
Thus, the suppression effect is very prominent in this situation, and
there is effectively complete suppression of the synchrotron radiation
for small values of $n$ (see also Fig.\ \ref{fig:liq}). In terms of
wavelength, the suppression is felt for wavelengths that are
comparable or longer than the mean distance between the
particles. Comparing (\ref{eq:gn-gas}) and (\ref{eq:gn-liq}) we
conclude that as the order becomes more manifest (temperature
decreases) the suppression effect becomes more prominent, but the
onset of suppression is shifted to longer wavelengths.

As crystallization takes place, long--range order effects become
important. We consider in the following the simplest crystal, namely
the linear chain. To describe the situation we assume a distribution
function that corresponds to a thermal distribution of small
displacements around the crystalline state with only nearest--neighbor
interactions taken into account for simplicity:
\begin{equation}
  P(\theta_1 , \ldots, \theta_N) =
  c_3 \exp \left[ - \eta \sum_{j=1}^{N}
    (\varphi_{j+1} - \varphi_j)^2 \right].
\end{equation}
Here $c_3$ is a normalization constant, $\eta \equiv (q^2 N^2)/(16
\pi^3 \epsilon_0 d k_{\rm B} T)$, $T$ is the temperature and 
$d \equiv 2 \pi \rho / N$ is the mean distance between the
charges. The small displacements $\varphi_j$ are defined as follows:
\begin{eqnarray}
  \varphi_1
  & \equiv &
  \theta_1,
  \\
  \varphi_j
  & \equiv &
  (\theta_j - \theta_1) - (j-1) \, d_{\theta} \; ,
    \; \; \; j = 2, 3, \ldots , N \ ,
  \\
  \varphi_{N+1}
  & \equiv &
  \varphi_1.
\end{eqnarray}
The exponent $\eta$ can be rewritten as
\begin{equation}
  \eta =
  \left( \frac{N}{2 \pi} \right)^2 \cdot
    \left( \frac{\rm typical \ potential \ energy}
                {\rm typical \ kinetc \ energy} \right)
  = \left( \frac{N}{2 \pi} \right)^2 \Gamma ,
\end{equation}
where $\Gamma$ is the plasma parameter in one dimension \cite{HG95}
\begin{equation}
  \Gamma = {q^2\over 4\pi\epsilon_0 d k_{\rm B} T}.
  \label{eq:PP}
\end{equation}
We note that the assumption of only nearest--neighbor interactions is
not severe since for small displacements the interaction with the
$n$'th neighbor reduces as $1/n^3$.

\begin{figure}[tp]

  \begin{center}
    \leavevmode
    \psfig{figure=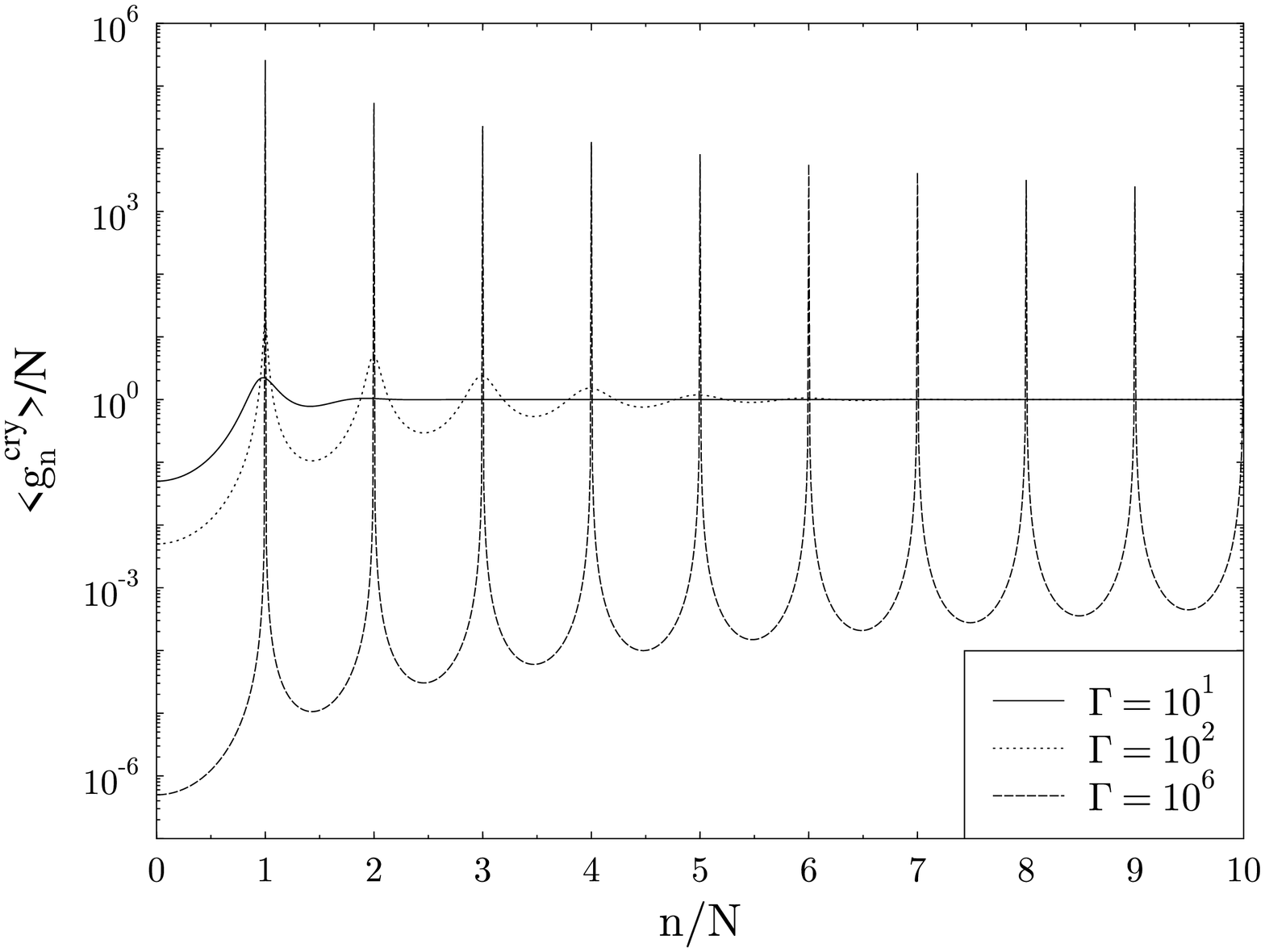,width=16cm}
  \end{center}

  \def\cap:cry{The form factor for the crystalline state of the beam,
  described by (\protect\ref{eq:ff-cry}). We considered the cases
  $N=10^6, \Gamma = 10^1, 10^2, 10^6$.}

  \caption{\cap:cry}

  \label{fig:cry}
\end{figure}

In this case it is easier to obtain the form factor directly, without
explicitly calculating the two--point correlator. A straightforward but
lengthy calculation gives the following result for the form factor
\begin{equation}
  \langle g_n^{\rm cry} \rangle =
  N + 2 \sum_{l=1}^{N-1} (N-l)
  \cos \left( \frac{2 \pi n l}{N} \right)
    \exp \left[ - \frac{n^2 \pi^2 l (N-l)}{\Gamma N^3} \right].
  \label{eq:ff-cry}
\end{equation}
In order to obtain the above result we assumed $\Gamma \gg 1$, i.e.\ a
cold beam. This is a necessary condition for crystallization.  In
order to interpret this result, we consider two limiting cases. If the
maximal exponent in (\ref{eq:ff-cry}) (as a function of $l$), given by
$n^2 \pi^2 / (4 \Gamma N)$, is much smaller than 1, we can replace the
exponential in (\ref{eq:ff-cry}) with $1$ and get
\begin{equation}
  \langle g_n^{\rm cry} \rangle \approx
  \cases{
    N^2 \ , &$N$ divides $n$, \cr
    0   \ , &otherwise.   \cr}
  \label{eq:ff-frozen}
\end{equation}
This means that for very cold crystals there is a total suppression of
the radiation for all harmonics, except the ones that are divisible by
the number of particles $N$. For these special harmonics we get total
constructive interference. The suppression of the leading harmonics
results in an enormous reduction of the {\em total} power emitted by
the synchrotron radiation (see Sec.\ \ref{sec:total-power}). In case
crystallized electron beams can be produced, this effect gives rise to
the possibility of significantly reducing the synchrotron radiation,
currently the main limitation for circular electron accelerators.  We
mention in passing that the result (\ref{eq:ff-frozen}) can also be
obtained directly from calculating the form factor for a completely
frozen crystal \cite{PB98,Sch46,Jac75}. The other limit of
(\ref{eq:ff-cry}) is for the first exponential factor ($l=1$) to be
already small, such that only the first term needs to be
considered. That is, for $n^2 \pi^2 / (\Gamma N^2) \gg 1$, we obtain
\begin{equation}
  \langle g_n^{\rm cry} \rangle \approx
  N + 2 N \cos \left( \frac{2 \pi n}{N} \right)
    \exp \left( - \frac{n^2 \pi^2}{\Gamma N^2} \right) ,
  \label{eq:ff-ripples}
\end{equation}
which describes small ``ripples'' over the incoherent radiation, with
decaying amplitude that has oscillations with period $N$. In Fig.\
\ref{fig:cry} we plot the numerically computed form factor
(\ref{eq:ff-cry}) for the specific cases $N = 10^6, \Gamma = 10^1,
10^2, 10^6$. For $\Gamma = 10^6$ we see a series of sharp peaks
located at $n/N=1,2,\ldots\,$. This is expected since in this case
$\Gamma$ is very large and thus (\ref{eq:ff-frozen}) holds
approximately for the range of $n$ shown in Fig.\ \ref{fig:cry}. For
smaller values of $\Gamma$ we observe a transition from sharp peaks to
decaying ripples. Even in the case $\Gamma = 10^6$ the sharp peaks
will eventually die away.

The depression of $g_n^{\rm cry}$ at $n \approx 1$ can be computed
analytically. Expanding the exponential factor in (\ref{eq:ff-cry}) to
first order in $1/\Gamma$ we obtain
\begin{equation}
  {1\over N}\,
  g_1^{\rm cry}\approx {1\over 2\Gamma}.
  \label{eq:g1cry}
\end{equation}
This is in perfect agreement with the results displayed in Fig.\
\ref{fig:cry}.

The above results concerning the crystalline state indicate that the
plasma parameter $\Gamma$ can be determined from the form factor of
the synchrotron radiation (provided $N$ is known). This defines a
useful {\em diagnostic} tool for measuring the temperature of the
crystal. We also conclude that crystalline beams can be {\em applied}
to selectively suppress and enhance harmonics of the radiation,
achieving up to total suppression ($g_n = 0$) or total constructive
interference ($g_n = N^2$).

To summarize this section, we have shown that the synchrotron
radiation and its modifications with respect to the incoherent state
are strongly connected with the physical state of the beam. The form
factor reflects the important scales and can be used to diagnose the
state of the beam (``gas'', ``liquid'', ``solid'') as well as its
temperature.

\section{Modifications for bunched beams}
\label{sec:bunched}

Experimentally it is sometimes useful to work with bunched beams in
which the particles occupy only a small fraction of the ring. Thus we
consider in this section the modifications of the above theory for
bunched beams. These modifications are straightforward. It turns out
that only the lowest harmonics (up to $n \approx 2 \pi/$(bunch angular
length)) are affected. Qualitatively this can be understood by
examining equation (\ref{eq:ff-prob}), since the bunching will be felt
only for values of $n$ such that $n ( \theta_j - \theta_{j'} )
\lesssim 2 \pi$. This yields the above estimate. In the following
we detail the theory quantitatively.

We start with the gaseous phase and consider a narrow bunch of $N$
particles with an (effective) angular width $\sigma \equiv 2 \pi / Q,
Q \gg 1$. In order to be specific we shall assume that the shape of
the bunch is a Gaussian, and that the (one--point) charge density is
given by equation (\ref{eq:p1-gaussian}) above. In the absence of
correlations, the two--point correlation function of the Gaussian bunch
reads
\begin{equation}
  \tilde{R}_2^{\rm GB}(\Delta) =
  \frac{N(N-1)}{\sqrt{\pi \sigma^2}}
  \sum_{m = -\infty}^{+\infty} \exp \left[
    - \frac{(\Delta + 2 \pi m)^2}{4 \sigma^2} \right]  ,
\end{equation}
from which we calculate the form factor (\ref{eq:ff-gb}).  In order to
include the hard--core repulsion between the charges, we operate as in
the coasting case and modify $\tilde{R}_2^{\rm GB}$ with a narrow dip
\begin{equation}
  \tilde{R}_2^{\rm gas, bunch} (\Delta) =
  c_4 \tilde{R}^{\rm GB}_2(\Delta)
  \left[ 1 - \exp \left( - \frac{\Delta^2}{2 a^2} \right) \right].
  \label{eq:r2-gas-bunch}
\end{equation}
As before, $c_4 \approx 1$ to leading order in $N$. When calculating
the form factor, the first term in the brackets gives
(\ref{eq:ff-gb}). For the second term, we can use $\tilde{R}_2^{\rm
GB} \approx 1$ since we assumed $a \ll
\sigma$. Hence, we obtain
\begin{eqnarray}
  \langle g_n^{\rm gas, bunch} \rangle
  & = &
  N - \frac{N^2 a}{\sqrt{2 \pi}} \exp \left( - \frac{n^2 a^2 }{2} \right)
  + N(N-1) \exp (-n^2 \sigma^2)
  \nonumber
  \\
  & = &
  \langle g_n^{\rm gas} \rangle + N(N-1) \exp (-n^2 \sigma^2) \ .
  \label{eq:gn-gas-bunch}
\end{eqnarray}
That is, the form factor of the gaseous coasting beam contains an
additional enhancement feature for low harmonics, $n \lesssim Q$. This
is suggestive, because of the scale separation between the length of
the bunch and the hard--core scale, $a \ll \sigma$. For the liquid
phase a similar analysis applies. We need to replace the term
$\tilde{R}_2^{0}$ in (\ref{eq:r2-liq}) with $\tilde{R}^{\rm GB}_2$,
and similar considerations will lead to the conclusion that we get the
same type of enhancement of the low harmonics due to bunching
\begin{equation}
  \langle g_n^{\rm liq, bunch} \rangle =
  \langle g_n^{\rm liq} \rangle
  + N(N-1) \exp (-n^2 \sigma^2).
\end{equation}

\begin{figure}[tp]

  \begin{center}
    \leavevmode
    \psfig{figure=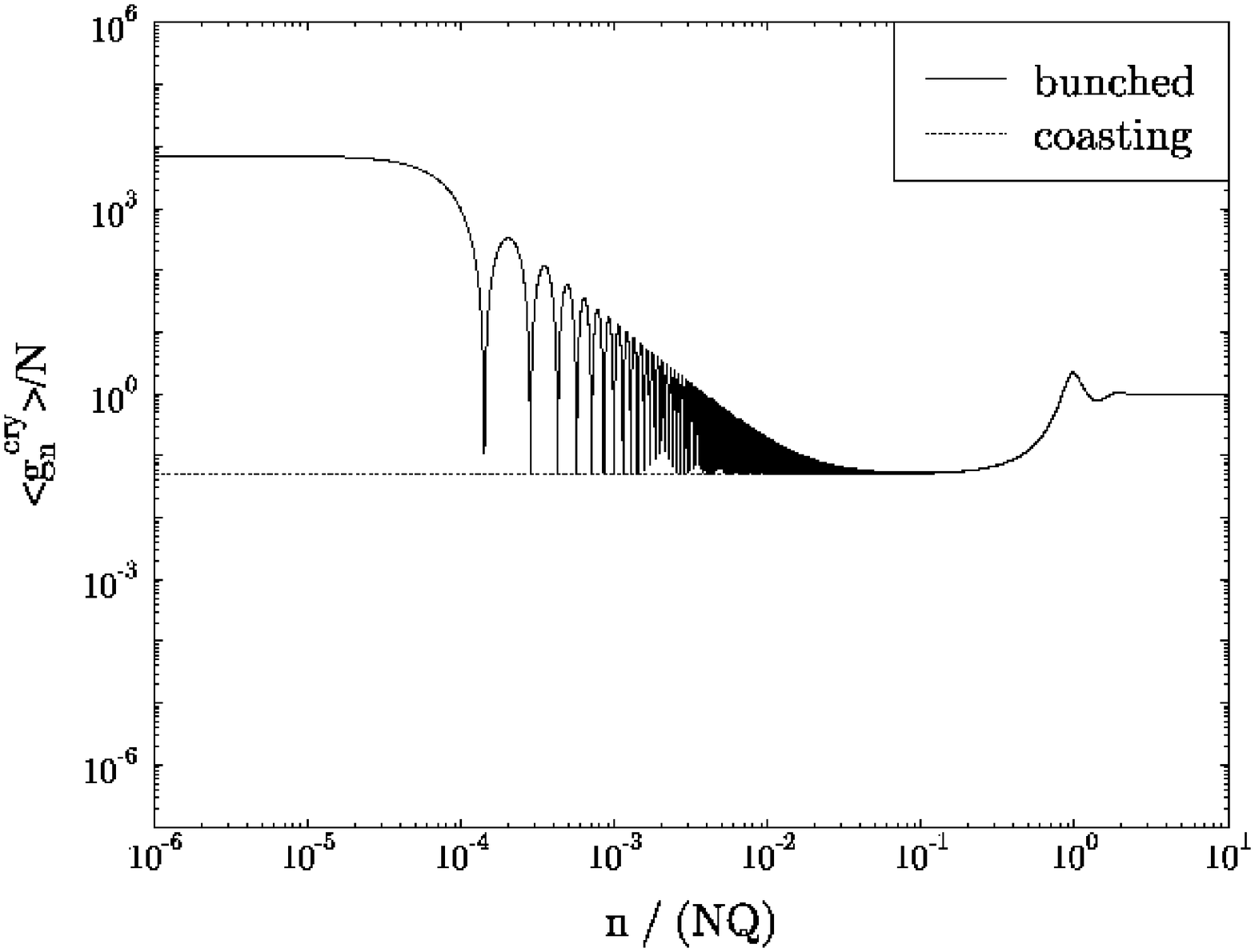,height=8cm}
    \psfig{figure=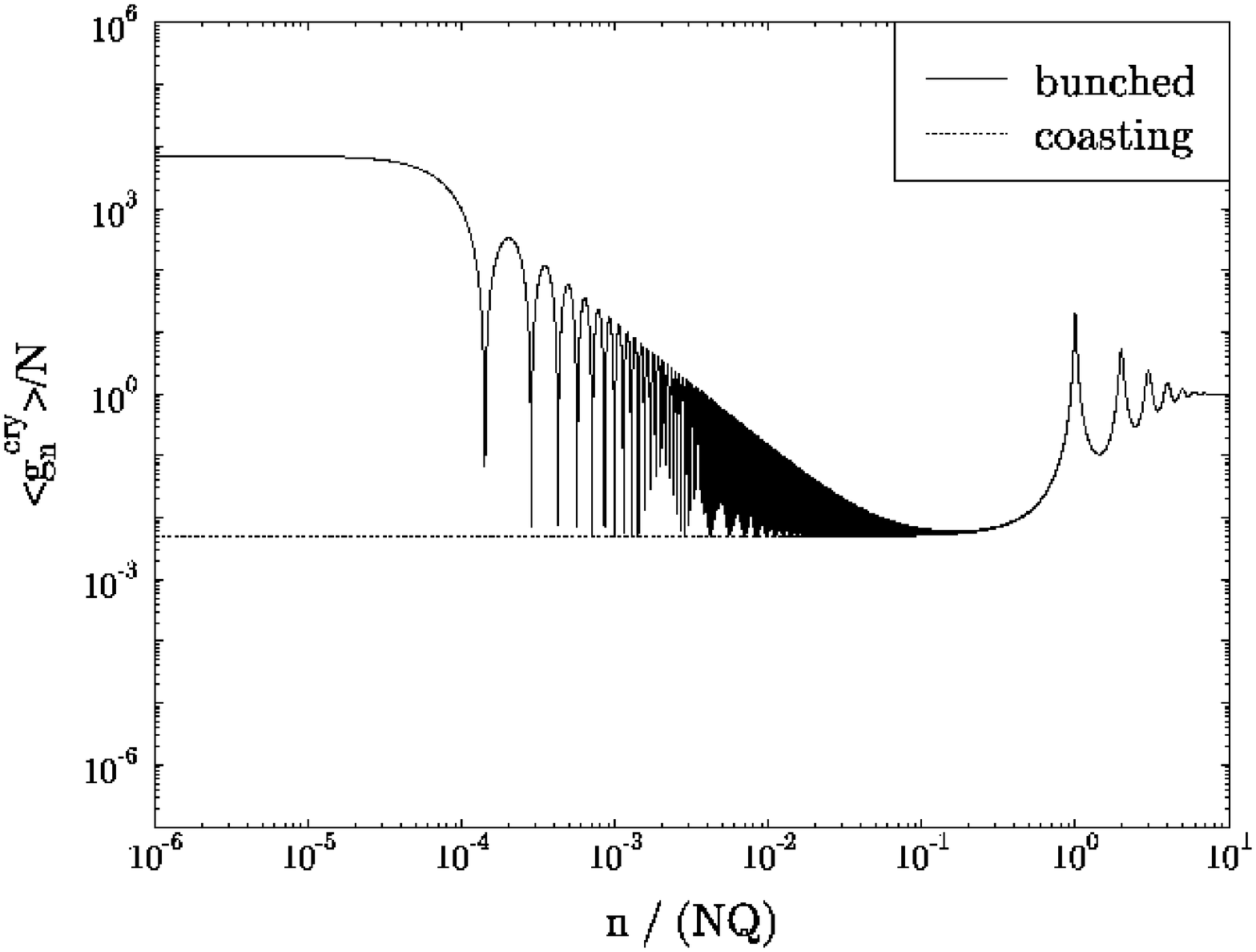,height=8cm}
  \end{center}

  \def\cap:cry-bunch{The form factor for the crystalline state of a
  bunched beam, described by (\protect\ref{eq:ff-cry-bunch}). We
  considered the cases $N=5 \sqrt{2} \times 10^3, Q = 100 \sqrt{2},
  \Gamma = 10^1$ (upper plot), $\Gamma = 10^2$ (lower plot).}

  \caption{\cap:cry-bunch}

  \label{fig:cry-bunch}
\end{figure}

For the crystalline state (linear chain) with finite temperature we
model the bunch by adding two stationary charges at both ends of the
bunch. These charges do not radiate and serve only for confinement. To
make the calculations tractable, we assume only nearest--neighbor
interactions. Lengthy but straightforward calculation yields the form
factor
\begin{equation}
  \langle g_n^{\rm cry, bunch} \rangle =
  N + 2 \sum_{l=1}^{N-1} (N-l) \cos(n l d_{\theta})
    \exp \left[ - \frac{n^2 d_{\theta}^2 l (N+1-l)}{4 \Gamma (N+1)}
        \right].
  \label{eq:ff-cry-bunch}
\end{equation}
For small values of $n$ we replace the exponents by $1$ and
obtain
\begin{equation}
  \langle g_n^{\rm cry, bunch} \rangle \approx
  \frac{\sin^2 ( \pi n / Q )}{\sin^2 [ \pi n / (Q N) ]}.
  \label{eq:ff-frozen-bunch}
\end{equation}
This is the form factor of a frozen linear crystalline bunch. In
particular, it exhibits an enhancement for the low harmonics $n
\lesssim Q$. If $n$ is so large that only the first term is
significant, we essentially recover the result (\ref{eq:ff-ripples}).
Results for the case $N = 5 \sqrt{2} \times 10^3, Q = 100 \sqrt{2}$
and $\Gamma = 10^1, 10^2$ are shown in Fig.\ \ref{fig:cry-bunch}. The
parameters were chosen such that $d_{\theta}$ is the same as for the
coasting case. We observe that significant enhancement indeed occurs
for the lower harmonics, which is essentially independent of the
temperature as suggested by (\ref{eq:ff-frozen-bunch}). Otherwise, the
form factor (normalized by the number of charges) is the same as for
the coasting case.

To summarize this section, we investigated the modifications that
result from the bunching of the particle beam.  In all cases we found
that a significant enhancement occurs for the low harmonics $n
\lesssim Q$. Otherwise we get qualitatively the same results as for a
coasting beam.

\section{Total power}
\label{sec:total-power}

In Secs.\ \ref{sec:form-factor} and \ref{sec:bunched} we concentrated
on a discussion of the form factor $g_n$ of the beam.  We showed that
important information on the thermodynamic state of the beam is
already contained in $g_n$.  The total emitted power, however, the
subject of this section, depends on the interplay between $g_n$ and
the partial power levels $I_n^{(1)}$ of a single radiating charge (see
Eq.\ (\ref{eq:totpow})). The total power $I^{(1)}$ of a single
radiating charge is given by
\begin{equation}
  I^{(1)} = 
  \sum_{n=1}^{\infty}\, I_n^{(1)} =
    {q^2 c \over 6 \pi \epsilon_0 \rho^2} \, \beta^4 \gamma^4 \, .
  \label{eq:tp}
\end{equation}
This result agrees with Larmor's well-known formula for the total
radiated power of a single charge in the nonrelativistic limit
\cite{Jac75}. We introduce the parameter $s=\beta\gamma$. It
characterizes the three relativistic regimes important for the
discussion in this paper: Nonrelativistic ($s\ll 1$), relativistic ($s
\approx 1$) and ultra-relativistic ($s\gg 1$). With the help of the
total power (\ref{eq:tp}) we define the normalized power levels
\begin{equation}
  {\tilde I}^{(1)}_n \equiv 
  {I^{(1)}_n\over I^{(1)}}.
\end{equation}
Since the purpose of this section is to discuss suppression effects in
the total emitted synchrotron--radiation power, we define the
suppression factor
\begin{equation}
  \alpha(N, \beta) \equiv
  {I^{(N)}\over N\, I^{(1)}}={1\over N}\,
    \sum_{n=1}^{\infty}\, g_n\, {\tilde I}^{(1)}_n \, .
  \label{eq:alpha-def}
\end{equation}
In the case of $N$ incoherently radiating charges we have
$\alpha(N,\beta)=1$.  A suppression effect corresponds to
$\alpha(N,\beta)<1$.  Enhancement of synchrotron radiation corresponds
to $\alpha(N,\beta)>1$.

\begin{figure}[p]

  \begin{center}
    \begin{tabular}{c}
      \leavevmode
      \psfig{figure=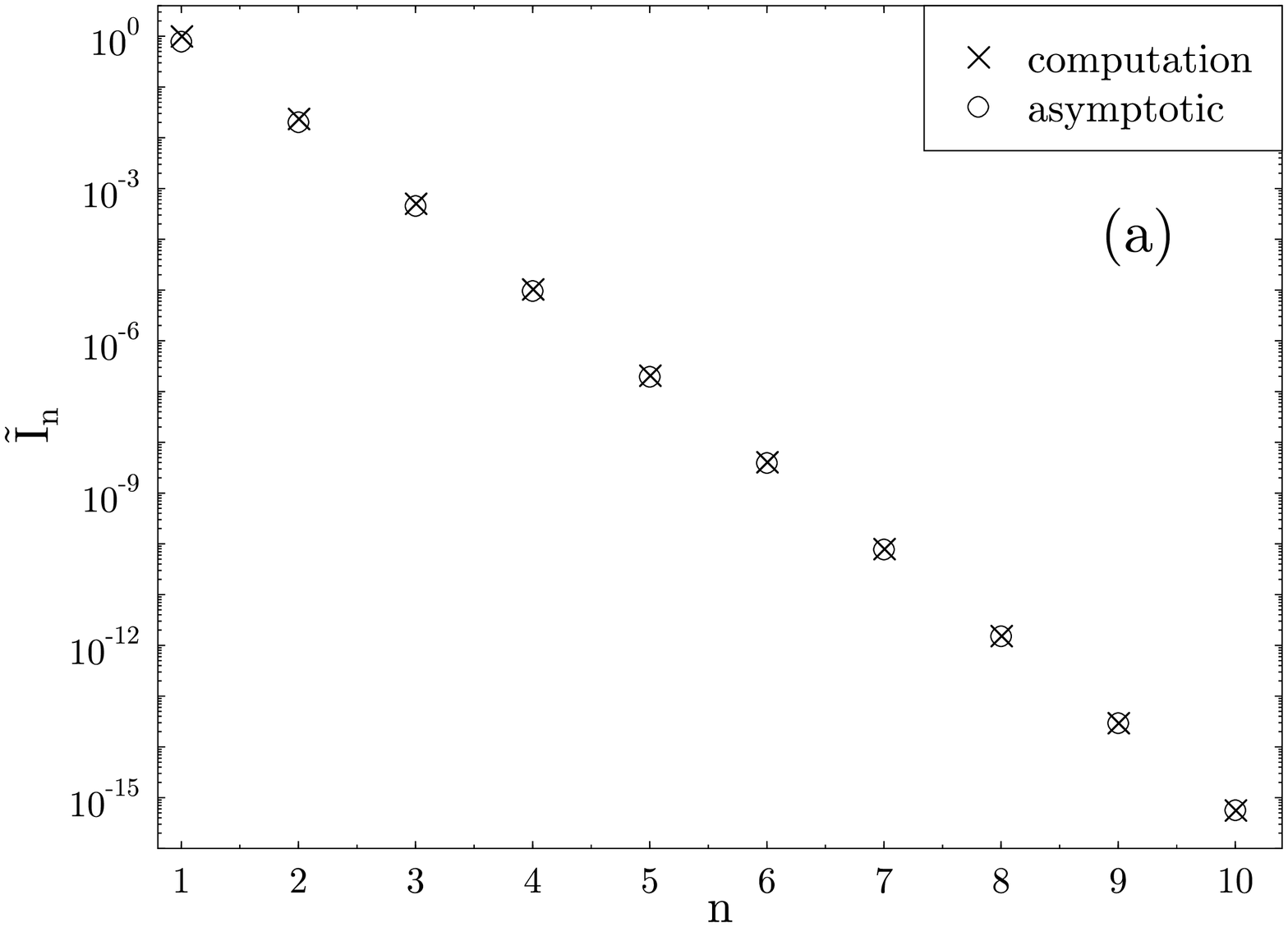,height=6cm} \\
      \psfig{figure=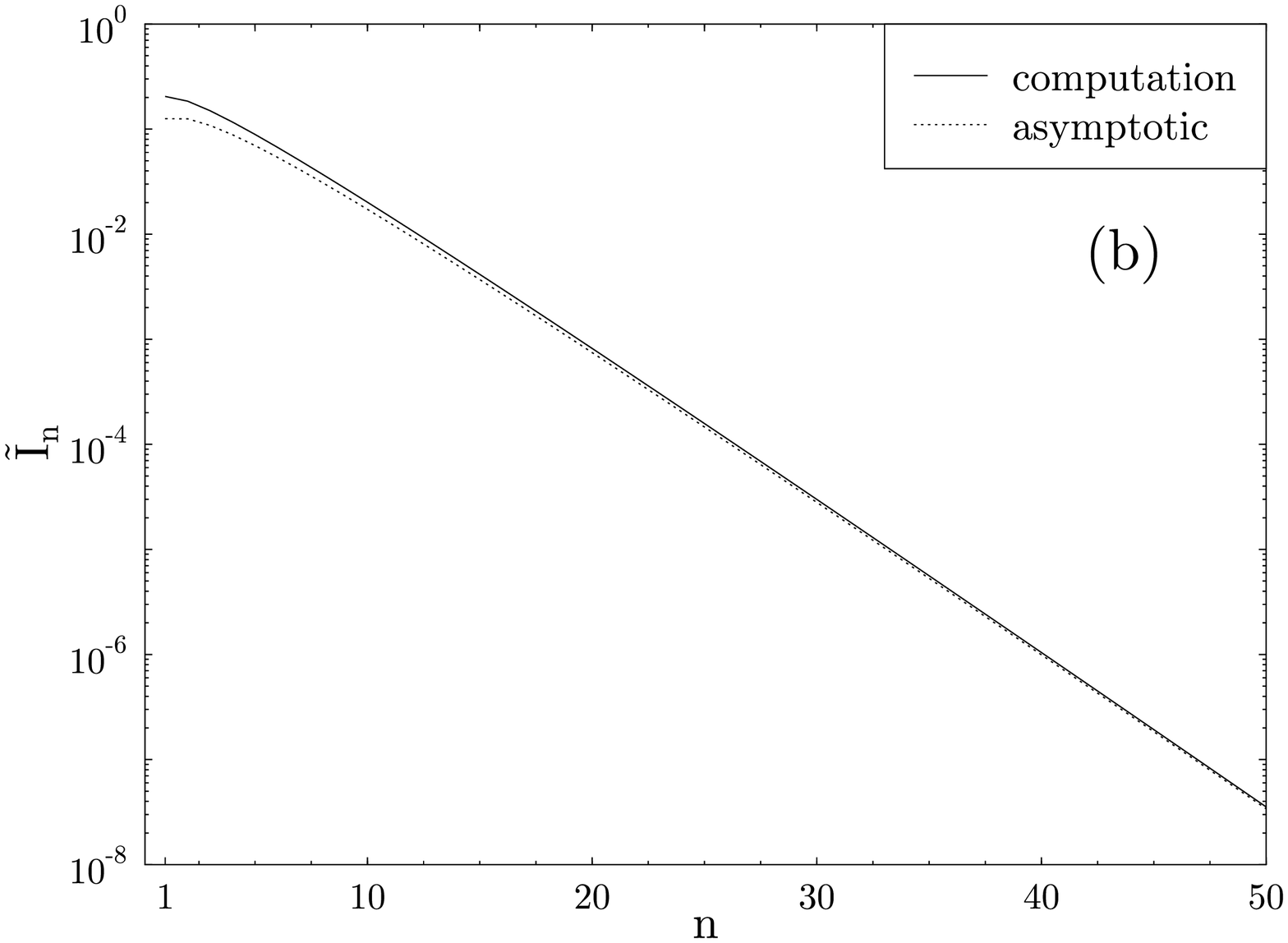,height=6cm}  \\
      \psfig{figure=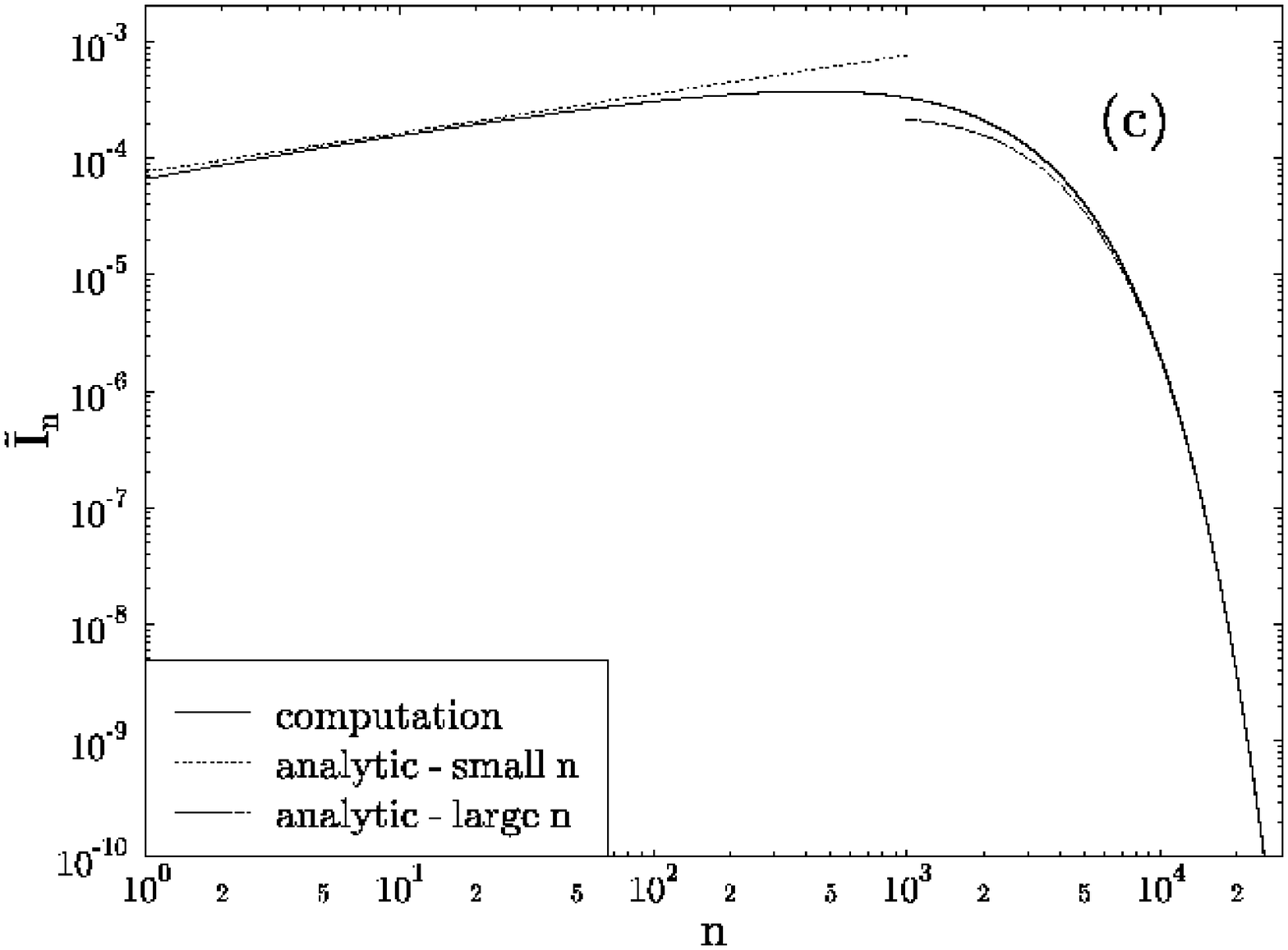,height=6cm}
    \end{tabular}
  \end{center}

  \def\cap:in{The normalized partial powers ${\tilde I}_n^{(1)}$ as a
  function of $n$ for (a) $s=0.1$, (b) $s=1$ (b) and (c) $s=10$. The
  analytical results (\protect\ref{eq:nrapprox}),
  (\protect\ref{eq:rapprox}), (\protect\ref{eq:ana1}) and
  (\protect\ref{eq:ana2}) are also shown in the respective panels.  }

  \caption{\cap:in}

  \label{fig:in}
\end{figure}

The behavior of ${\tilde I}_n^{(1)}$ as a function of $n$ is the key
for understanding the suppression effect of the total emitted
synchrotron power. It is qualitatively different in the three
relativistic regimes (see Fig.\ \ref{fig:in}). For $s\ll 1$ we have
$\beta\ll 1$ and ${\tilde I}_n^{(1)}$ decays exponentially in $n$.
This is illustrated in Fig.\ \ref{fig:in}(a). It shows ${\tilde
I}_n^{(1)}$ as a function of $n$ for $s=0.1$. Expanding
(\ref{eq:I1nexact}) to leading order in $\beta$ we obtain
\begin{equation}
  {\tilde I}_n^{(1)} \approx 
  {3(n+1)n^{2n+1}\over (2n+1)(2n)!}\,
  \beta^{2n-2} \, , \ \ \ 
  \beta \ll 1.
\end{equation}
We verify that ${\tilde I}_1^{(1)} \approx 1$ in this limit. Using
Stirling's formula we obtain
\begin{equation}
  {\tilde I}_n^{(1)} \approx 
  {3(n+1)\sqrt{n}\over
  2(2n+1)\sqrt{\pi}\beta^2}\,
  \left({e\beta\over 2}\right)^{2n} ,
  \ \ \ \beta \ll 1,\ \ n \gg 1 \ ,
  \label{eq:nrapprox}
\end{equation}
which proves the exponential decay of ${\tilde I}_n^{(1)}$ for large
$n$. The result (\ref{eq:nrapprox}) is also shown in Fig.\
\ref{fig:in}(a). The exponential decay for large $n$ persists in the case 
$s \approx 1$, albeit with a much smaller decay constant.  This is
illustrated in Fig.\ \ref{fig:in}(b). In this case we also have an
analytical approximation. It is given by \cite{LL79}
\begin{equation}
  {\tilde I}_n^{(1)} \approx
    {3\sqrt{n}\over 4 \sqrt{\pi} \beta^2 \gamma^{9\over 2}}
    \left( \frac{\beta \gamma {\rm e}^{\frac{1}{\gamma}}}
                {1 + \gamma} \right)^{2n}
    \; , \; \; 1 \lesssim \gamma \, , \, n \gg \gamma^3 \; .
  \label{eq:rapprox}
\end{equation}
The analytical approximation (\ref{eq:rapprox}) is shown as the dashed
line in Fig.\ \ref{fig:in}(b). It describes the numerical data very
well. The same figure also shows that a qualitative change with
respect to the nonrelativistic case (Fig.\ \ref{fig:in}(a)) occurs
only for small $n$ where ${\tilde I}_n^{(1)}$ starts with a near--zero
slope.  In the ultra-relativistic case ($s\gg 1$) the behavior of
${\tilde I}_n^{(1)}$ changes qualitatively.  For small $n$ it shows an
initial power-law increase according to \cite{LL79}
\begin{equation}
  {\tilde I}_n^{(1)}\approx 
  0.78 \, \gamma^{-4} n^{1/3} \, , \; 
  \gamma \ll 1 \, , \; 1 \ll n \ll \gamma^3 \, .
  \label{eq:ana1}
\end{equation}
At $n \approx 0.29 \gamma^3$ it reaches a maximum and then decays
exponentially according to \cite{LL79}
\begin{equation}
  {\tilde I}_n^{(1)} \approx
  {3\sqrt{n} \over 4 \sqrt{\pi} \gamma^{9 \over 2}}
    \exp \left( - \frac{2n}{3 \gamma^3} \right)
  \; , \; \; 
  \gamma \gg 1 \, , n \gg \gamma^3 \; .
  \label{eq:ana2}
\end{equation}
This behavior is illustrated in Fig.\ \ref{fig:in}(c) for the case
$s=10$ (full line). The analytical results (\ref{eq:ana1}) and
(\ref{eq:ana2}) (dashed lines) are also shown in Fig.\
\ref{fig:in}(c). They compares well with the data in the appropriate
limits. We now show that the behavior of ${\tilde I}_n^{(1)}$ in
conjunction with the behavior of $g_n$ leads to substantial
suppression of synchrotron radiation for cold beams.

\begin{figure}[tp]

  \begin{center}
    \leavevmode
    \begin{tabular}{c}
      \psfig{figure=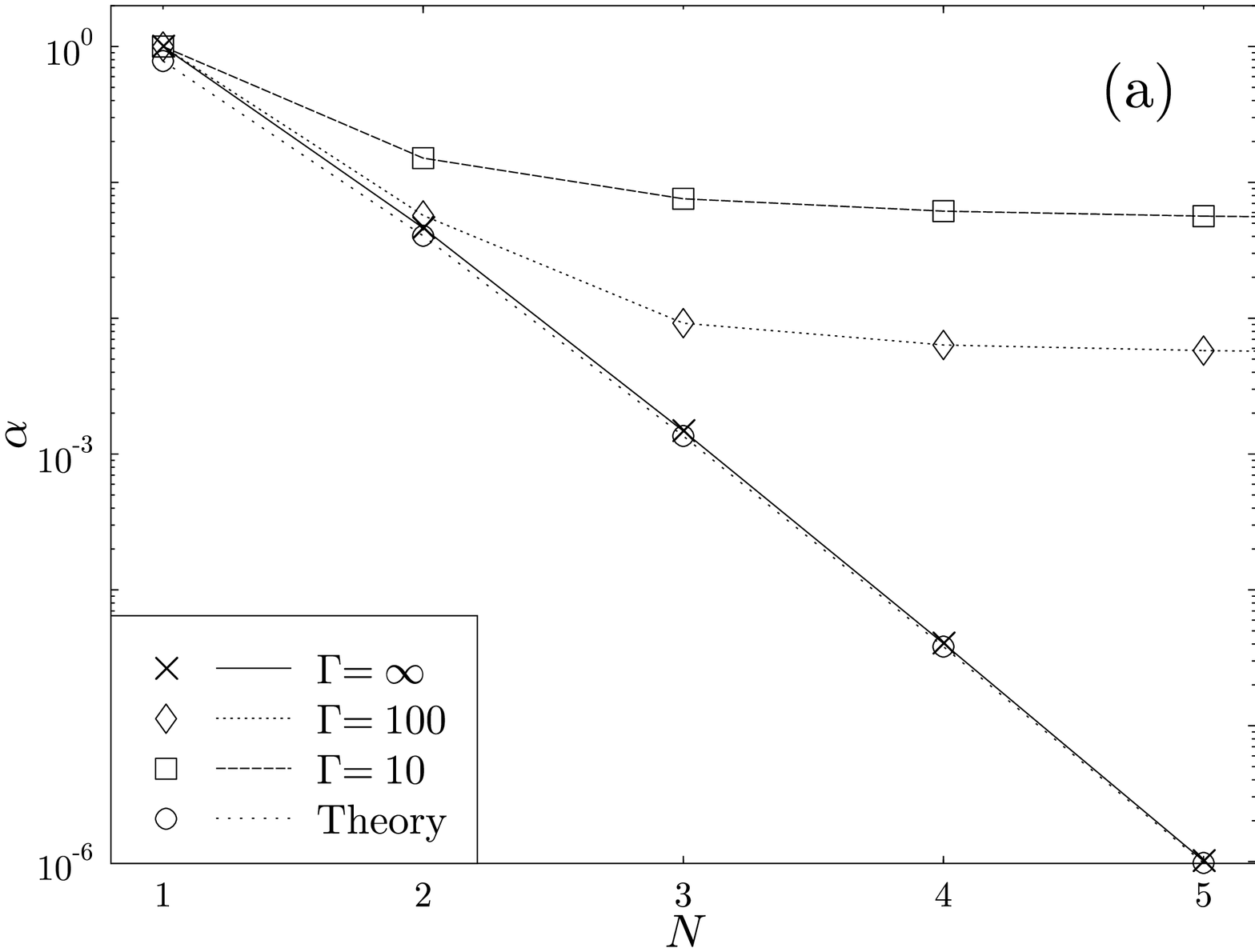,height=6cm} \\
      \psfig{figure=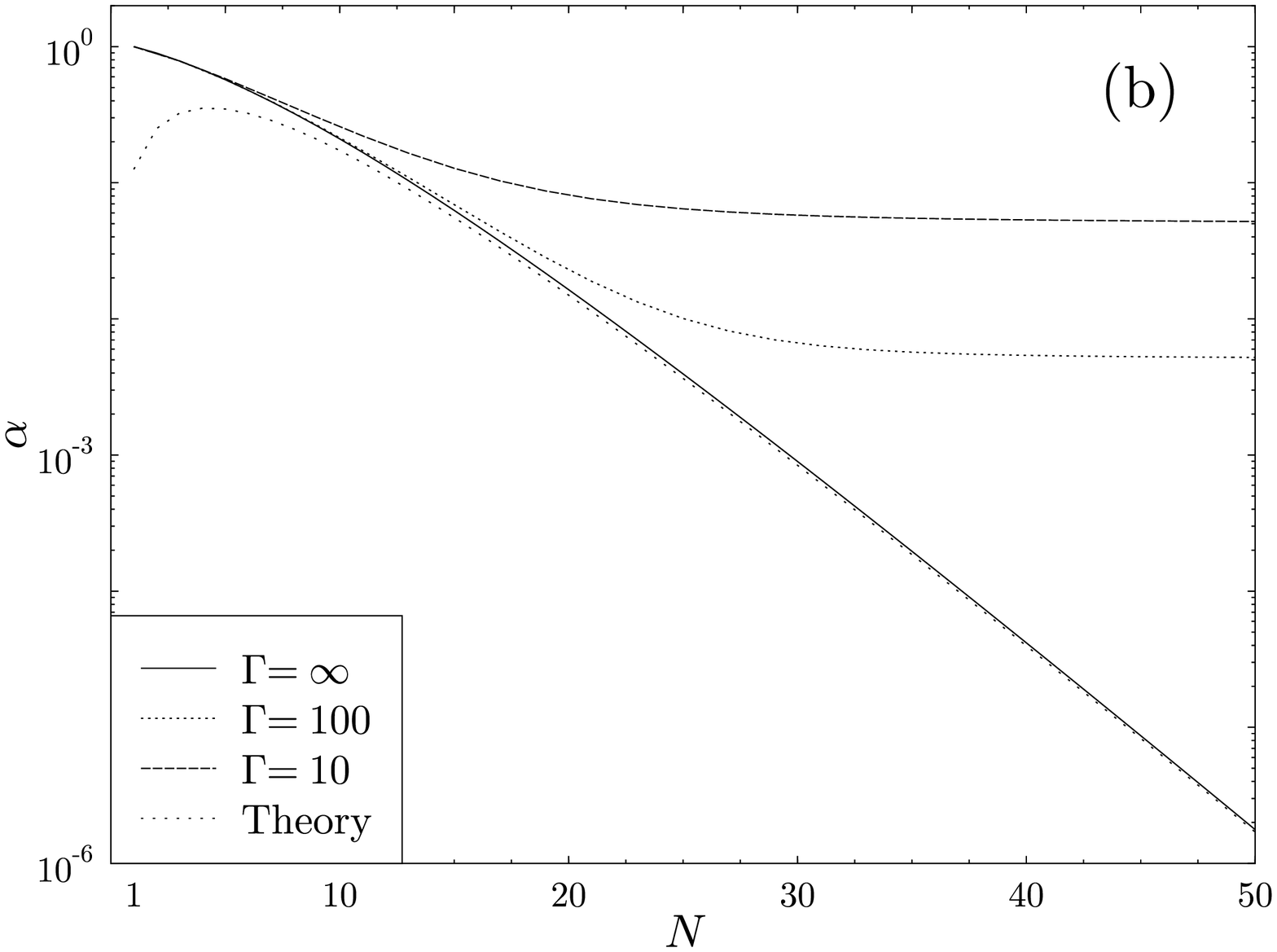,height=6cm} \\
      \psfig{figure=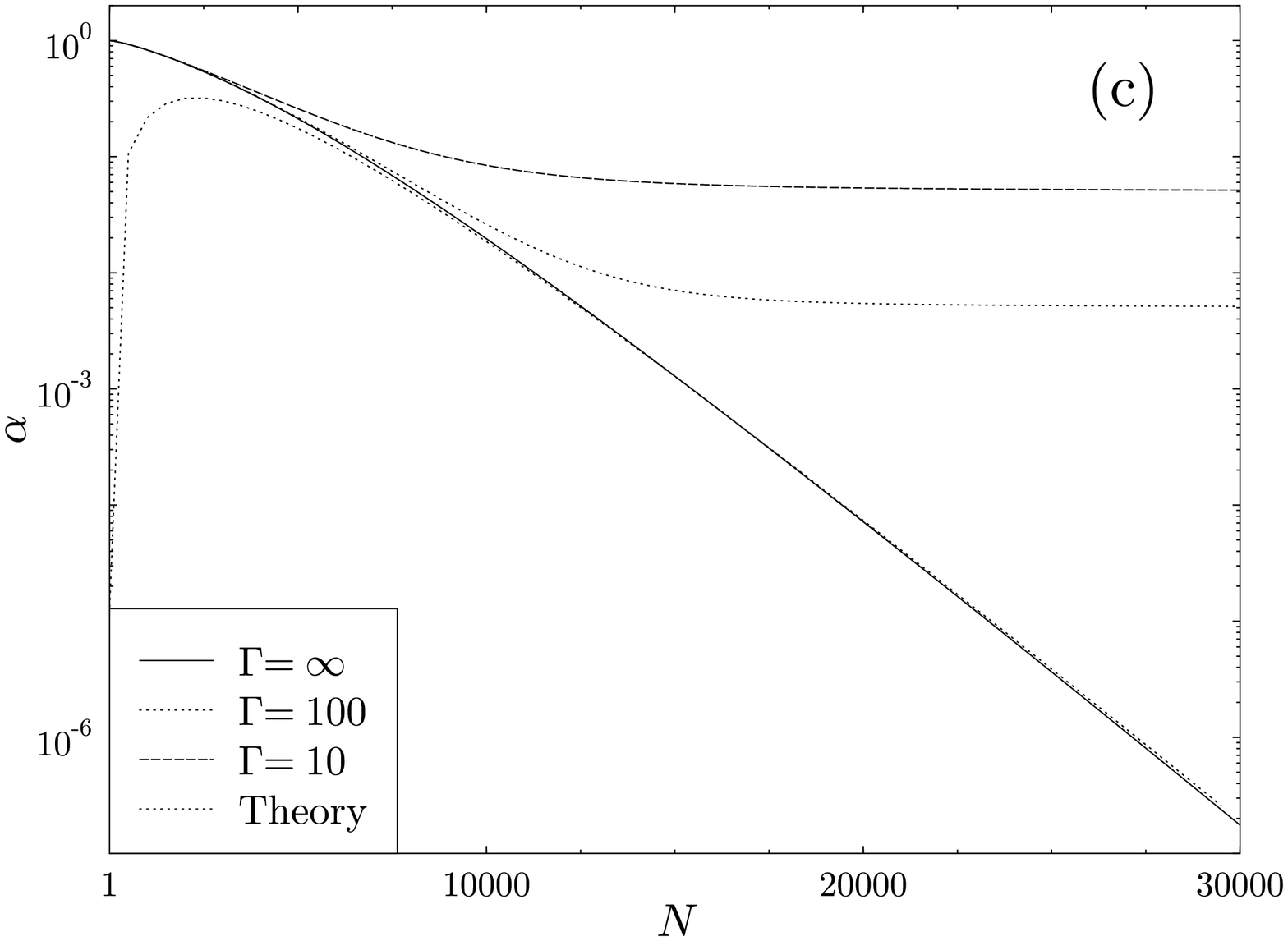,height=6cm}
    \end{tabular}
  \end{center}

  \def\cap:supp{Suppression factors for the crystallized chain for
  three different plasma parameters ($\Gamma=10,100,\infty$) in the
  three relativistic regimes: (a) $s=0.1$, (b) $s=1$ and (c) $s=10$.
  The theoretical curves correspond to equations
  (\protect\ref{eq:nrsupp}), (\protect\ref{eq:sup-int}) and
  (\protect\ref{eq:sup-ult}).  }

  \caption{\cap:supp}

  \label{fig:supp}
\end{figure}

We first discuss the case of a coasting crystallized linear chain at
$T=0$. It consists of $N$ equi-spaced particles according to
$\theta_j=2\pi j/N$, $j=1,2,\ldots ,N$. For $g_n$ we have the result
(\ref{eq:ff-frozen}).  For the suppression factor $\alpha$ we obtain
in this case
\begin{equation}
  \alpha(N,\beta) = 
  N \, \sum_{m=1}^{\infty} \, {\tilde I}^{(1)}_{m\cdot N} \, .
  \label{eq:supp}
\end{equation}
We saw above that independently of $s$ the normalized partial powers
${\tilde I}^{(1)}_n$ always decay exponentially for large enough $n$.
Thus, there is always an $N_0$ such that
\begin{equation}
  \alpha \approx 
  N \tilde{I}_{N}
  \label{eq:alpha-large-N}
\end{equation}
to a very good approximation.  Therefore $\alpha$ is exponentially
small for $N > N_0$. In other words: For large enough particle number
we obtain exponential suppression of synchrotron radiation
independently of the relativistic regime of the beam. This result is
illustrated in Fig.\ \ref{fig:supp} ($\Gamma = \infty$ case). It shows
the suppression factor for $s=0.1$, 1 and 10 as a function of the
particle number $N$. In all three cases we indeed obtain exponential
suppression as predicted from the structure of (\ref{eq:supp}).
Using Eq.\ (\ref{eq:alpha-large-N}) and the above expressions for
$\tilde{I}_n$ in the relevant relativistic regimes, we obtain explicit
analytical formulae for $\alpha(N, \beta)$:
\begin{eqnarray}
  \alpha(N, \beta) 
  & \approx &
  {3 N^{\frac{3}{2}} \over 4\sqrt{\pi}\beta^2}
  \left( {{\rm e} \beta\over 2}\right) ^{2N} \, ,
  \; s \ll 1 \, , \; N \gg 1 \; ,
  \label{eq:nrsupp}
  \\
  \alpha(N, \beta)
  & \approx &
  \frac{3 N^{\frac{3}{2}}}{4 \sqrt{\pi} \beta^2 \gamma^{\frac{9}{2}}}
  \left( \frac{\beta \gamma {\rm e}^{\frac{1}{\gamma}}}
              {1 + \gamma} \right)^{2N} \, ,
  \; s \lesssim 1 \, , \; N \gg \gamma^3 \; ,
  \label{eq:sup-int}
  \\
  \alpha(N, \beta)
  & \approx &
  \frac{3 N^{\frac{3}{2}}}{4 \sqrt{\pi} \beta^2 \gamma^{\frac{9}{2}}}
  \exp \left( - \frac{2N}{3 \gamma^3} \right) \, ,
  \; s \gg 1 \, , \; N \gg \gamma^3 \; .
  \label{eq:sup-ult}
\end{eqnarray}
Fig.\ \ref{fig:supp} shows that the analytical formulae are very good
approximations of the numerical data in their respective ranges of
validity.

Next we consider the linear chain at finite temperature.  In this case
the form factor (\ref{eq:ff-cry}) applies.  Because of the structure
of (\ref{eq:ff-cry}) and the asymptotic exponential decay of ${\tilde
I}^{(1)}_n$ for large $n$ we can compute the asymptotic behavior of
$\alpha(N, \beta)$ for large $N$. Using (\ref{eq:g1cry}) we obtain
\begin{equation}
  \alpha(N, \beta) = 
  {1 \over N}\sum_{n=1}^{\infty}g_n^{\rm cry}
    {\tilde I}^{(1)}_n  \approx {1\over N} g_1^{\rm cry}
    \sum_{n=1}^{\infty} {\tilde I}^{(1)}_n =
  g_1^{\rm cry}/N \approx 
  {1\over 2\Gamma}, \ \ \ N \gg \gamma^3 \, .
\end{equation}
Thus, for large $N$ and in all three relativistic regimes, the
asymptotic suppression is independent of $N$ and saturates at
$\alpha=1/(2\Gamma)$. This behavior is illustrated in Fig.\
\ref{fig:supp} which shows the suppression factor for $\Gamma=10$, 100
and $\infty$ for all three values of $s$ considered.  The onset of
saturation in the vicinity of some $N=N_c$ is physically clear because
of the following reason. Finite $\Gamma$ corresponds to a finite
temperature which furthermore corresponds to a finite correlation
length of the particles in the linear chain. But since the suppression
of the synchrotron radiation is a coherent process it is intuitively
clear that no further suppression can be achieved once the total
particle number exceeds the correlation length. Consequently the
suppression effect has to saturate.

In Sec.\ \ref{sec:form-factor} we pointed out that measuring the depth
of the correlation hole in $g_n^{\rm cry}$ for small values of $n$
defines an experimental method for measuring the plasma parameter of
the beam.  Since the saturation value of $\alpha$ depends only on
$\Gamma$, measuring the suppression factor for large $N$ defines yet
another experimental procedure for measuring $\Gamma$.

The existence of a finite correlation length at finite temperature
provides an argument why it is not necessary to maintain coherence
over the whole circumference of the storage ring in order to observe
the suppression effect. It is enough to work with bunches whose length
is smaller or of the order of $N_c$ to observe suppression of
synchrotron radiation. From the mathematical point of view this is
also evident since we saw that for finite temperature only lower
harmonics of $g_n$ are affected by the bunching, hence $g_{m \cdot
N}^{\rm cry,bunch} \approx g_{m \cdot N}^{\rm cry}$, $m=1,2,\ldots$
for large enough $N$, and therefore we expect similar suppression as
for the coasting case. This result is very important for practical
applications of the suppression effect. It means that the coherence
does not have to be maintained over the whole extent of the ring,
which sometimes can amount to hundreds of meters and more. It is
enough to maintain the crystalline structure over small angular
distances (bunches) in order to exploit the suppression effect in
possible technical applications.

It is also clear by inspection of Fig.\ \ref{fig:in} and of Fig.\
\ref{fig:liq}(b) that for large enough $N$ substantial suppression of
synchrotron radiation can be achieved for liquid beams. This, again,
is important since modern electron coolers are close to providing a
liquid beam of electrons.  Thus it may soon be possible to check our
theory with the help of liquid electron beams.

\section{Discussion, summary and conclusions}
\label{sec:summary}

The suppression of the radiation of geometrically ordered charges was
first noticed by J. J. Thomson \cite{Tho08}. He employed this effect
for motivating the stability of atoms, which, according to classical
theory, should radiate and decay. Suppression of synchrotron radiation
in the context of accelerators was first noted by
L. I. Schiff\cite{Sch46}. But in Schiff's time a mechanism for
establishing the order in a beam of charged particles was not
available. Only recently, with progress in the cooling of beams by
electrons and lasers is it possible to envision the production of
crystallized beams whose synchrotron radiation is exponentially small.
It should be born in mind, however, that synchrotron radiation is not
very important for heavy ion beams that can easily be cooled with
electrons and lasers. Dramatic effects are expected to occur only for
crystallized electrons where the synchrotron radiation is orders of
magnitude stronger. The draw-back is that electrons cannot be cooled
directly with lasers. We hope, however, that this paper will stimulate
experimentalists to develop cooling schemes for electron beams. One
possibility would be to use sympathetic cooling of electrons with a
beam of heavy ions that can be cooled by lasers.

The paper discusses various forms of ordered beams that may occur in
practice: Gaseous, liquid and crystalline, coasting and bunched.  It
is pointed out that the suppression effect occurs on two levels: In
the form factor of the beam and in the total radiated power. While the
modifications in the form factor may be used as a diagnostic tool for
inferring the thermodynamic state of the beam, the suppression of the
total power may eventually lead to the construction of small--sized
cyclic electron accelerators.

\section*{Acknowledgements}

HP is grateful for a MINERVA fellowship. RB is grateful for financial
support by the Deutsche Forschungsgemeinschaft (SFB 276).



\end{document}